# Resonant magneto-optic rotation for magnetometry using autonomous frequency stabilization

S. Pradhan[1, 2,*], S. Mishra[1], R. Behera[1], Poornima[1,2], and K. Dasgupta[1,2]
[1]Laser and Plasma Technology Division, Bhabha Atomic Research Centre, Mumbai-85, India
[2]Homi Bhabha National Institute, Department of Atomic Energy, Mumbai-85, India
[*]Corresponding author: spradhan@barc.gov.in

The operation of a high sensitive atomic magnetometer using resonant elliptically polarized light is demonstrated. The experimental geometry allows autonomous frequency stabilization of the laser, thereby offers compact operation of the overall device. The magnetometry is based on measurement of the zero magnetic field resonance in degenerate two level system using polarimetric detection and has a preliminary sensitivity of <10 pT/Hz$^{1/2}$ @ 1 Hz.

The prospect of atomic magnetometers is reflected in their compact size and operation near room temperature apart from high sensitivity. These possibilities has given them an unique advantages for a variety of futuristic applications like magneto-cardiography, magneto-encephalography, stellar & interstellar field mapping, underground & underwater surveys etc [1-4]. The requirements for these specialized applications are very specific in terms of sensitivity, dynamic range, and sizes. For example, the biomedical application desires miniaturized sensor head where as overall size & weight of the device is of paramount importance for space application. For either biomedical or space application, the all optical magnetometer has an edge over RF magnetometer [5] due to possible reduction in sizes.

The all optical magnetometer has two variant, one operating on degenerate two level systems and the other relying on quantum interference among distinct hyperfine levels. The measurement based on degenerate two level systems primarily operates as NMOR (non-linear magneto-optic rotation) or SERF (spin exchange relaxation free) magnetometer [6, 7]. The preparation of the atomic system for these two techniques are different, however relies on similar detection scheme. Recently, Shah and Romalis have circumvented the requirement of the additional pump beam in the orthogonal direction to the probe beam for the operation of the SERF magnetometer by the use of an elliptically polarized light without compromising much on the sensitivity [8]. This magnetometer was operated at ~45 GHz away from the atomic resonance and the elliptically polarized light does the dual job of spin polarization and measurement of the magnetic field.

The all optical magnetometer working on quantum interference between hyperfine states can be operated either in high field regime, where the separation between the coherent population trapping (CPT) states is a measure of magnetic field or in the low field regime utilizing the separation between the convoluted transmission and polarization rotation signal arising due to quantum interference [9-12]. The later techniques provide a unique possibility of simultaneous operation of atomic clock and atomic magnetometer apart from operational flexibility for magnetometry [12].

Here we have envisaged a new experimental geometry for measurement of the magnetic field utilizing zero magnetic field resonance of the degenerate two level systems. It is based on the analyzer parallel to the polarizer configuration against conventional 45$^0$ among them as has been utilized for either NMOR or SERF magnetometer. We use conventional frequency modulation spectroscopy to improve the signal to noise ratio. The schematic experimental set-up is shown in Fig.-1.

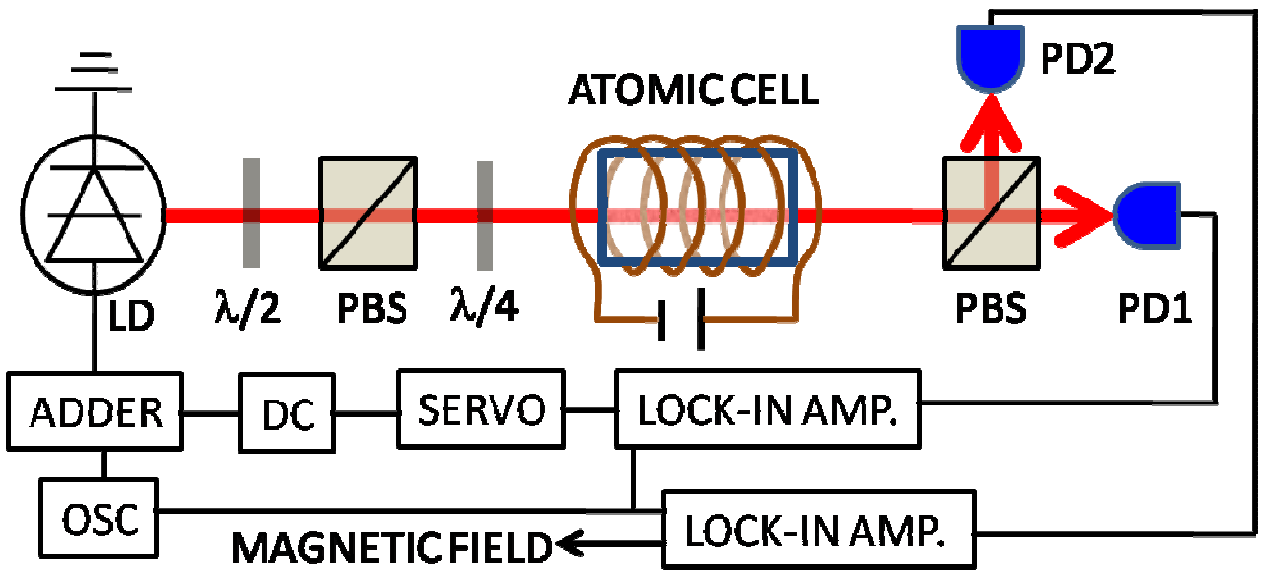

***Figure-1:*** *In the above experimental set-up, a VCSEL laser diode is frequency modulated by adding an oscillating current to the DC injection current. The desired elliptically polarized light is generated using a half wave plate, a polarizer and a quarter wave plate. The atomic sample kept in a temperature and magnetic field controlled environment, changes the polarization and power of the input beam near atomic resonance. The transmitted signal by the analyzer kept parallel to the polarizer is demodulated with a lock-in amplifier and is used to stabilize the laser frequency by controlling the DC injection current through a servo loop. The demodulated reflected signal directly provides the magnetic field.*

The Experiment is carried out with a vertical cavity surface emitting diode laser. The laser diode (LD) is stabilized at $24^0$ C using thermo electric cooler, transistor based sensor and a servo loop. The DC voltage used to drive the LD is added with an AC voltage @ ~6 kHz for frequency modulation spectroscopy. The diode laser beam polarization and intensity is controlled with a half wave plate, a polarizing beam splitter cube (PBS, named polarizer) and a quarter wave plate. The desired ellipticity in the laser beam is achieved by placing the quarter wave plate at $\sim 8^0$ with respect to the polarizer. The laser beam power is ~ 230 μW with knife edge width of ~4.5 mm at the interaction regime. The atomic cell contains Rb atoms in natural isotopic composition in presence of Neon buffer gas at 50 Torr. The diameter and the length of the cell is 25 mm. The atomic cell is heated up to $\sim 58^0$ C using a pair of resistive wire carrying current in opposite direction. The heating wires are enclosed in Mu-metal shield and DC current is used for the heater operation. A fixed current is passed through the heater coil throughout the experiment. The background magnetic field in the orthogonal direction to the laser propagation direction is attenuated using several layers of cylindrical Mu-metal shield. The magnetic field along the laser propagation direction and residual field in the orthogonal direction are controlled by three sets of Helmoltz magnetic coils in mutually orthogonal direction. All the magnets are calibrated with respect to the CPT resonances between different hyperfine levels in high field [10-12] and are used to bring three axis magnetic field close to zero (within few nT) at the interaction regime. It's worth mentioning that any other method, (like taking help of another magnetometer) can also be utilized for realization of the near zero field. The elliptically polarized light interacts with the atomic sample near zero magnetic field. The change in the laser intensity and polarization rotation is detected with an analyzer placed parallel to the polarizer and a pair of photodiodes (PD) shown in the Fig.-1. The transmitted light through the analyser is phase sensitively detected with a lock-in amplifier and is used to stabilize the laser frequency using a servo loop. The reflected light by the analyzer with autonomous frequency stabilized laser is phase sensitively detected for magnetometry.

The optical part of the experimental system including laser diode, thermoelectric cooler, optics, atomic cell, heater assembly, magnetic field shield, three axis magnetic field control and detectors are enclosed inside a box of size 28 cm × 8 cm × 8 cm. The whole optical component can be miniaturized using MEMS technology [9]. The electronics control and signal processing circuit are developed in our laboratory and enclosed in a box of size 30 cm × 20 cm × 10 cm. The whole experiment can be performed with or without supervision of a computer. It may be noted that, the same set-up is used for magnetic field calibration using CPT states from distinct hyperfine levels using an additional RF driver [10-12].

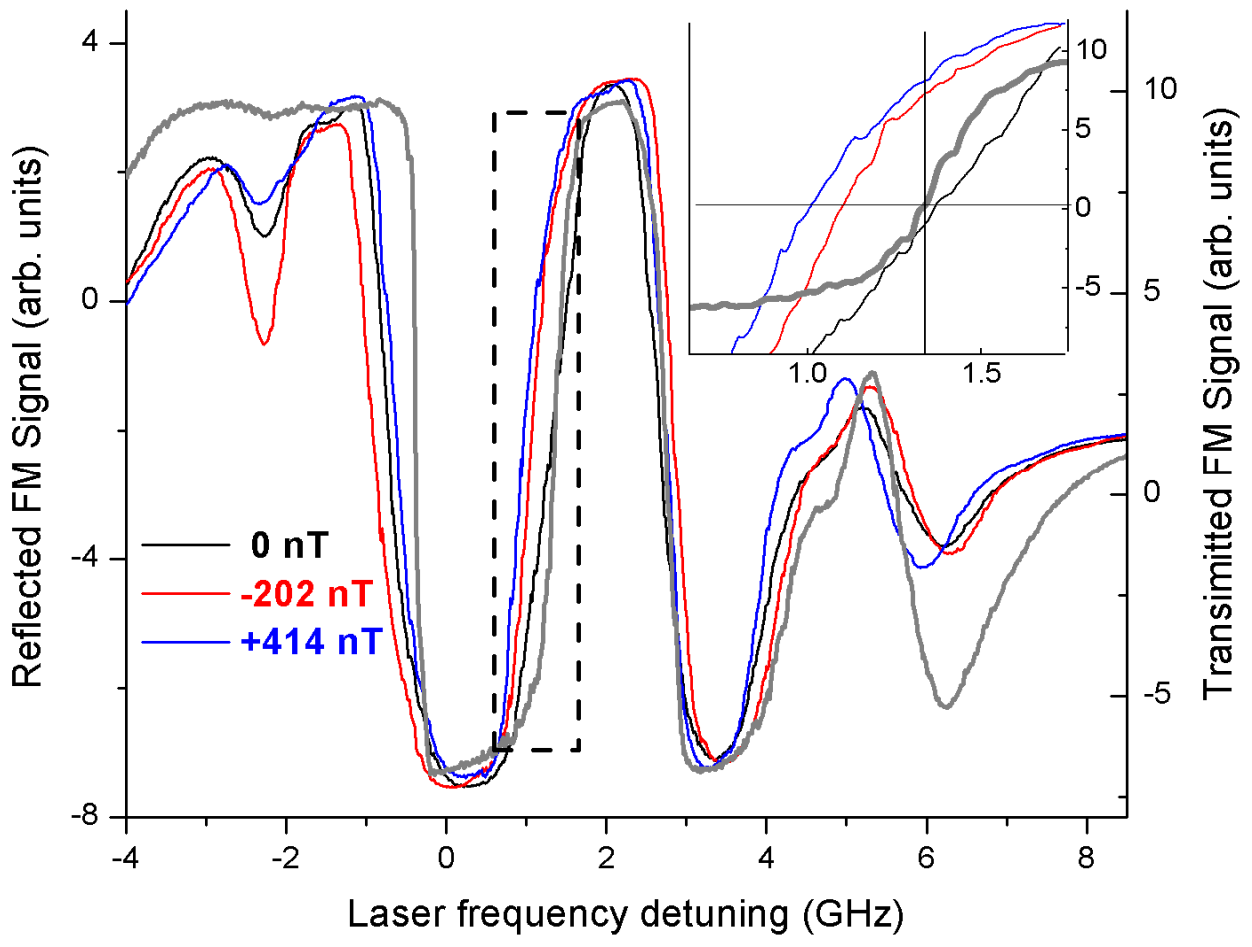

*Figure-2: The reflected signals by the analyzer as a function of laser detuning with respect to the Rb-85 F=3 to F'=3 hyperfine transition are shown for various magnetic field. The transmitted signal (gray line) is shown for ~0 nT and doesnot change significantly in the region marked by the dashed square. The insert shows the expanded portion of the dashed square. It is clear from the insert plot that reflected signal significantly varies with magnetic field (changes along the vertical solid line) for laser frequency locked to the zero crossing of the transmission signal. All signals are obtained using FM spectroscopy.*

The transmitted and the reflected light by the analyzer as a function of laser frequency are shown in figure-2. The signals correspond to the derivative of the actual profile due to implementation of FM spectroscopy. The illustrated signal profile provides a guideline for the location of the single photon detuning where magnetometry can be realized with best possible sensitivity. This is carried out by taking the transmission and polarization rotation signal for various magnetic fields along the laser propagation direction as a function of laser detuning. There are two important criteria for choosing the single photon detuning for magnetometry using the present technique. After matching the transmission signal for various magnetic field at the targeted single photon detuning, firstly the reflected signal should have largest possible variation with respect to change in the magnetic field and secondly the transmitted signal should have a large slope. The first criteria determine the sensitivity of the magnetometer where as the second one is important for stabilization of the laser frequency. One of the possibilities is shown in figure-2, where the transmission signals for various magnetic fields are matched in the region marked by the dashed square and is represented by gray plot taken at ~0 nT. The insert clearly shows that the reflected polarization rotation signal is very sensitive to applied magnetic field. In the present experiment we obtain maximum sensitivity of the magnetometer for laser locked to ~+1.3 GHz away from Rb-85 F=3 to F'=3 hyperfine transition and corresponds to the zero crossing of the transmitted FM signal shown in Fig.-2.

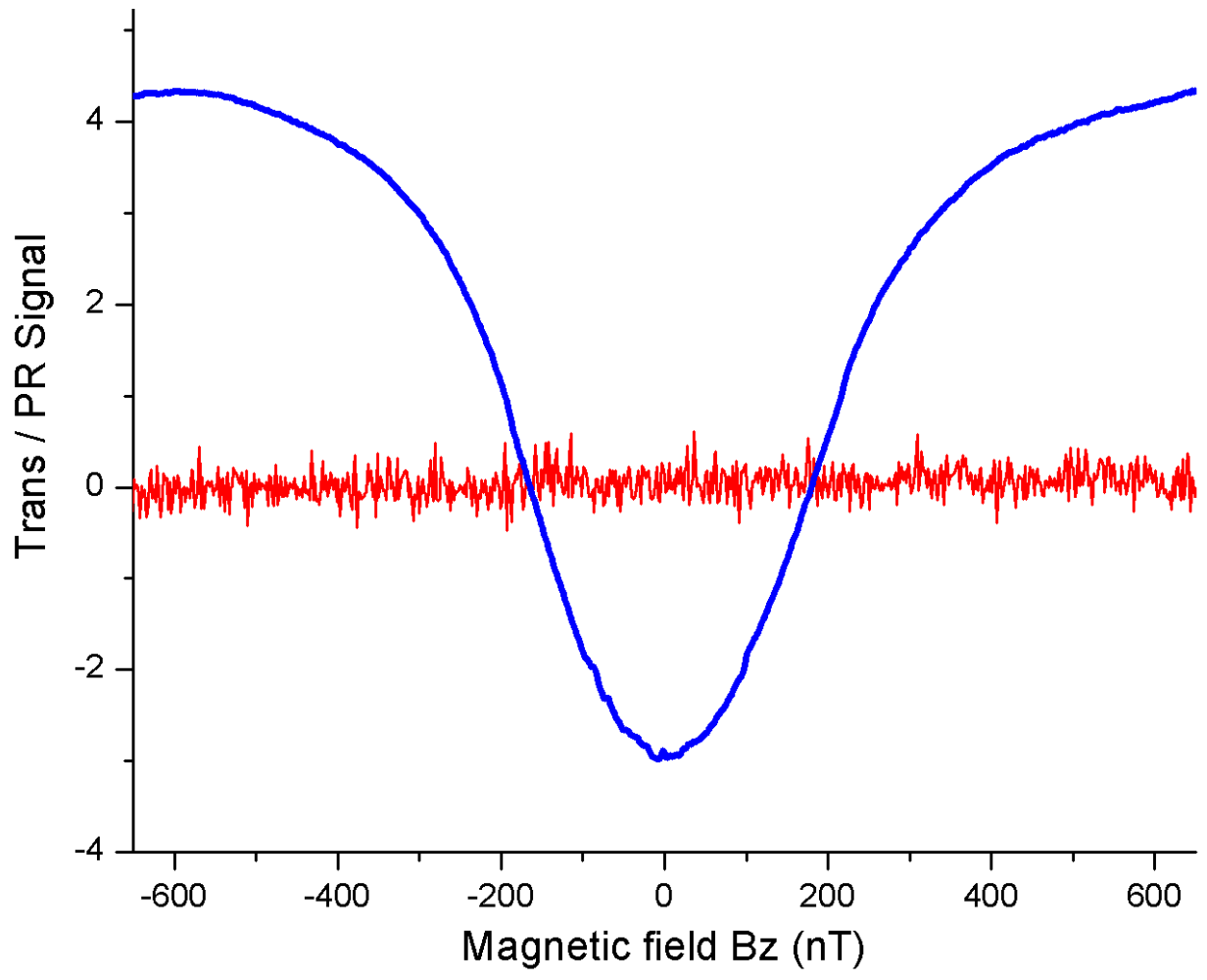

*Figure-3: The zero magnetic field resonance in the polarization rotation signal (blue) as a function of magnetic field along the laser propagation direction. The transmitted signal (red) of the analyzer in Fig.-1 is used to frequency stabilize the diode laser at its zero crossing as shown in Fig.-2. The width of the signal is related to the dynamic range and in combination with signal to noise ratio provide sensitivity of the device.*

The zero magnetic field resonance obtained with the present technique is shown in figure-3. Here the laser frequency is locked to the above mention zero crossing of the transmission signal of Fig.-2. The polarization rotation signal with this autonomous frequency stabilized laser is shown as the magnetic field is scanned across the zero field. The illustrated magnetic field resonance is originated from the Zeeman coherence in degenerate two level systems [6, 8, 13-15]. The linearly polarized light in sigma configuration can be viewed as a combination of $\sigma^+$ and $\sigma^-$ component, when its interaction with an atomic system is envisaged. The propagation of these two light field in the atomic system is accompanied by a differential phase shift or/and a differential absorption due to real and imaginary part of the complex refractive index respectively. These effects are popularly termed as Faraday effect or optical dichroism respectively. Consequently there is a polarization rotation or/and change in ellipticity of the light beam which can be measured using a polarimetric detection system. Introducing a small imbalance among $\sigma^+$ and $\sigma^-$ component, further induces a population imbalance resulting in enhanced polarization rotation (and/or dichroism). This elliptically polarized light can induce optical activity in the atomic system and is known as polarization self rotation [16]. The polarization self rotation can occur either in presence or in absence of magnetic field.

For an atomic system, the linearly polarized light with $\sigma^+$ and $\sigma^-$ component can simultaneously couple to ground and excited state Zeeman substates creating V or $\Lambda$ kind of coupled states. The Raman resonance for these coupled systems can be achieved near zero magnetic field even when a single laser beam is used, thereby establishing coherence among the Zeeman substates. The optical activities of the atomic system get significantly altered due to these CPT states, consequently changing the reflected signal from the analyzer. The reflection is further facilitated by using an elliptically polarized light where the $\sigma^+$ and $\sigma^-$ component are unequal. This has been extensively studied for coherence among the Zeeman substate corresponding to different hyperfine level [10,12]. Though the present observation is related to establishment of coherence among Zeeman sub state corresponding to the same hyperfine state, here also polarization rotation is facilitated by the unequal $\sigma^+$ and $\sigma^-$ components. In summary, it is the establishment of Zeeman coherence along with an elliptically polarized light that result in change in reflection of the light from the analyzer.

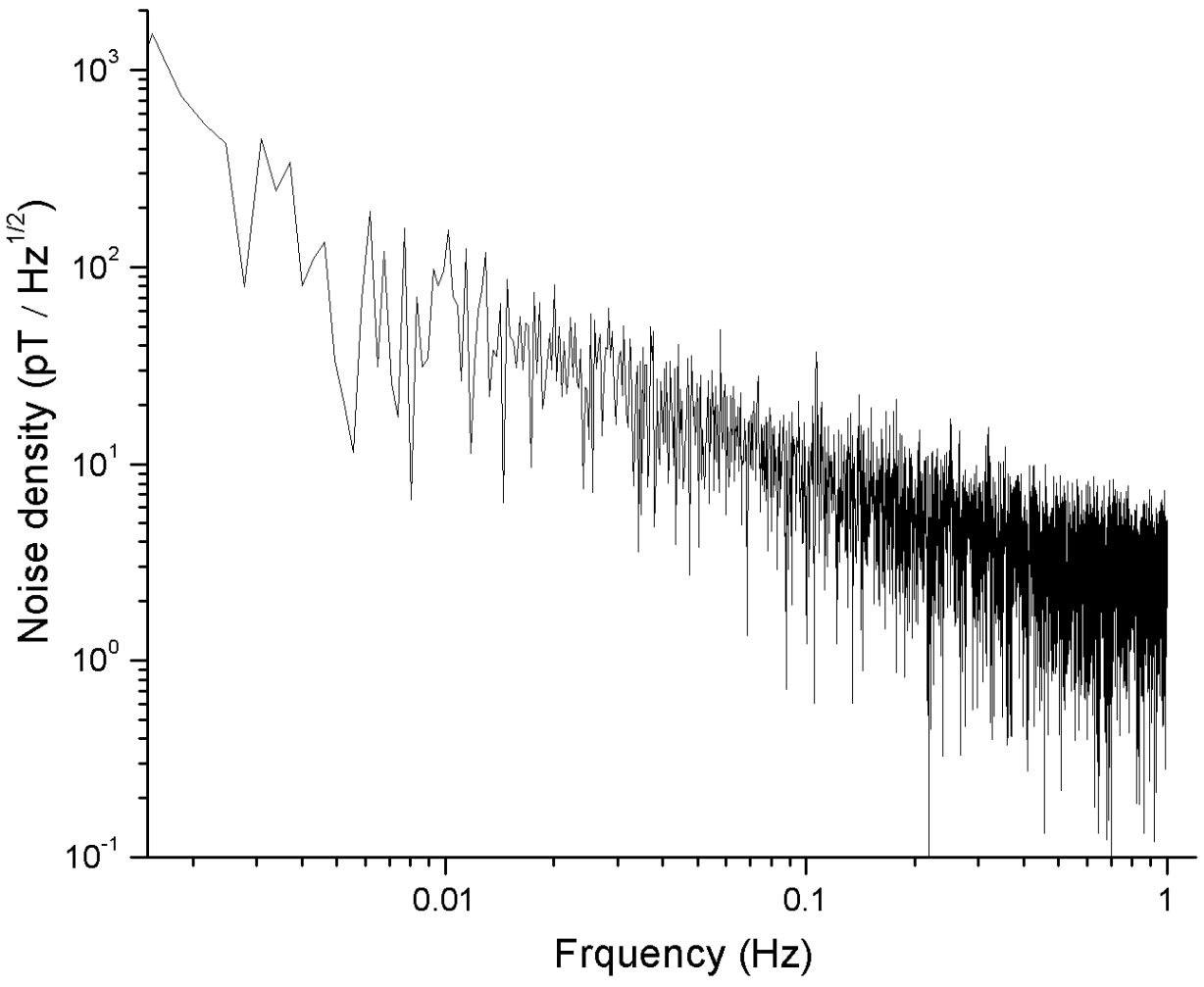

*Figure-4: Square root of power spectral density is shown as a function of frequency. The noise density is <10 pT/Hz$^{1/2}$ beyond 0.1 Hz.*

The illustrated magnetic resonance is central to the magnetic field measurement demonstrated here. Magnetic field can be measured directly using the polarization rotation signal shown in Fig.-3, or by applying a modulating magnetic field and generating the derivative of the signal for magnetometry. Here we have used the illustrated magnetic field resonance to calculate the sensitivity of the magnetometer. It is realized by applying a DC magnetic field of ~200 nT and taking a long time series data of the polarization rotation signal with the laser frequency locked to the transmitted signal. The square root of the Fourier transform of the auto correlator function provides the magnetic field noise as shown in figure-4. The magnetometer has a peak sensitivity of <10 pT/Hz$^{1/2}$ @ 1 Hz, which can be further reduced by improvising electronics circuit and optimizing the various experimental parameters.

**Conclusions:** An atomic magnetometer with a sensitivity of <10 pT/Hz$^{1/2}$ @ 1 Hz is demonstrated. The magnetometer is all optical and operated with a single elliptically polarized light field. The experimental geometry offer autonomous frequency stabilization of the laser frequency, thereby useful for application (e.g. space and others) where overall size of the device is important.

**Acknowledgements:** The authors are sincerely thankful to Dr. L.M. Gantayet, Director, BTDG for fruitful discussion and support during this work.